\documentstyle[epsfig,aps,prl,twocolumn]{revtex}
\newcommand{\mZ}{m_Z}
\newcommand{\mW}{m_W}
\newcommand {\kn}{K^-}

\newcommand {\Be}{{\cal{B}}_{e}}
\newcommand {\Bm}{{\cal{B}}_{\mu}}
\newcommand {\Bp}{{\cal{B}}_{\pi}}
\newcommand {\Bk}{{\cal{B}}_{K}}
\newcommand {\Bl}{{\cal{B}}_{\ell}}
\newcommand {\Bh}{{\cal{B}}_{h}}
\newcommand {\MTval}     {(1776.96^{+0.18+0.25}_{-0.21-0.17})\mathrm{MeV}}   
\newcommand {\TAUTval}   {(290.55     \pm 1.06)        \,\mathrm{fs}}    
\newcommand {\BRTEval}   {(17.786     \pm 0.072)       \,\%}             
\newcommand {\BRTMval}   {(17.356     \pm 0.064)       \,\%}             
\newcommand {\BRTPval}   {(11.01      \pm 0.11)        \,\%}             
\newcommand {\BRTKval}   {(0.692      \pm 0.028)       \,\%}             
\newcommand {\GF}        {G_{{F}}}

\newcommand {\GFMUval}   {(1.16639  \pm 0.00002) \times 10^{-5}\,\mathrm{GeV^{-2}}} 
\newcommand {\FPIVUDval} {(127.4 \pm 0.1)  \mathrm{MeV}}     %
\newcommand {\FKVUSval}  {(35.18 \pm 0.05) \mathrm{MeV}}     %
\newcommand {\MNUTvalA} {32}    
\newcommand {\MNUTvalB} {38}    
\newcommand {\FMIXvalA} {0.007} 
\newcommand {\FMIXvalB} {0.008} 
\begin{document}
%
%
%
\draft
\title{New constraints on the tau neutrino mass and fourth generation mixing}
%
%
\author{John Swain and Lucas Taylor}
\address{Department of Physics, Northeastern University, Boston, USA}
\date{\today}
\maketitle
\begin{abstract}
We present new constraints on the mass $m_{\nu_\tau}$ of the tau 
neutrino and its mixing with a fourth generation neutrino.
From an analysis of the partial widths of tau lepton decays
we obtain the following bounds at the 90\% confidence level:  
\mbox{$m_{\nu_\tau}<\MNUTvalA$\,MeV} and 
\mbox{$\sin^2\theta<\FMIXvalA$}, 
where $\theta$ describes the Cabibbo-like mixing of the third and 
fourth generation neutrinos.
\end{abstract}
%
%
\section{Introduction}
In a previous paper we derived constraints on the mass 
of the third generation neutrino $\nu_3$
and its mixing with a heavy fourth generation neutrino $\nu_4$~\cite{MNUTAU}.
In this paper we update this analysis using recent experimental measurements.
We determine significantly more stringent constraints on the mass $m_{\nu_3}$
and the Cabibbo-like mixing angle $\theta$, where 
the tau neutrino weak eigenstate is given by the
superposition of two mass eigenstates 
$|\nu_\tau\rangle = \cos\theta |\nu_3\rangle 
+ \sin\theta |\nu_4\rangle$.      
We compare the precise measurements of the $\tau$ partial 
widths for the following decays%
\footnote{Throughout this paper the charge-conjugate decays are also implied.
          We denote the branching ratios for these processes as
          $\Be, \Bm, \Bp, \Bk$ respectively;
          $\Bl$ denotes either $\Be$ or $\Bm$ while $\Bh$ denotes either $\Bp$ or $\Bk$.}
:
$\tau^-\rightarrow{e}^-\bar{\nu}_{{e}}\nu_\tau$,
$\tau^-\rightarrow\mu^-\bar{\nu}_\mu\nu_\tau$,
$\tau^-\rightarrow\pi^-\nu_\tau$, and
$\tau^-\rightarrow{K}^-\nu_\tau$,
with our theoretical predictions, as functions of 
$m_{\nu_3}$ and $\sin^2\theta$ to obtain upper limits on both 
these quantities. 
 
\section{Theoretical predictions}
The theoretical predictions for the branching fractions $\Bl$ for the 
decay $\tau^-\rightarrow\ell^-\bar{\nu}_{\ell}\nu_\tau (X_{\mathrm{EM}})$, with
$\ell^-={e}^-, \mu^-$ and $X_{\mathrm{EM}} = \gamma,~\gamma\gamma,~e^+e^-,\ldots$, 
are given by:
\begin{eqnarray}
  \Bl^{\mathrm{theory}}  
              & = & \left(\frac {\GF^2 m_\tau^5}{192\pi^3}\right)\tau_\tau                                         
                    \left( 1 -  8x - 12 x^2{\mathrm{ln}}x + 8 x^3 - x^4\right)      \nonumber \\
              &   & \times  \left[ \left( 
                                          1 - \frac{\alpha(m_\tau)}{2\pi} 
                                          \left( 
                                                 \pi^2 - \frac{25}{4} 
                                          \right) 
                            \right)  
                           \left(1 + \frac{3}{5} \frac{m_\tau^2}{\mW^2} 
                           \right) \right]  \nonumber  \\
             &   & \times                                
                    \left[ 1 - \sin^2\theta \right]
                    \left[ 1 - 8y(1-x)^3+\cdots\right]
\label{equ:blept}
\end{eqnarray}
where 
$\GF    = \GFMUval$ is the Fermi constant~\cite{PDG96SHORT};
$\tau_\tau = \TAUTval$ is the tau lifetime~\cite{LI97A};     
$m_\tau = \MTval$~\cite{MTAUBESNEW} is the tau mass, determined
by BES from the $\tau^+\tau^-$ production rate
near threshold which has no dependence on the 
tau neutrino mass;
and 
$x=m_\ell^2/m_\tau^2$.
The first term in brackets allows for radiative 
corrections\cite{BERMAN58A,KINOSHITA59A,SIRLIN78A,MARCIANO88A}, 
where $\alpha(m_\tau)\simeq 1/133.3$ is the QED coupling constant~\cite{MARCIANO88A} 
and $\mW = 80.400 \pm 0.075$\,GeV is the W mass~\cite{PIC97}.
The second term in brackets describes mixing with a fourth generation neutrino
which, being kinematically forbidden, causes a suppression of the decay rate.
The third term in brackets  parametrises the suppression 
due to a non-zero mass of $\nu_3$, where $y=m_{\nu_3}^2 / m_\tau^2$
and the ellipsis denotes negligible higher order 
terms~\cite{MNUTAU}.                                

The branching fractions for the decays $\tau^-\rightarrow{h}^-\nu_\tau$, 
with ${h}=\pi/{K}$, are given by
\begin{eqnarray}  
  \Bh^{\mathrm{theory}}  
             &  = &    \left(\frac {\GF^2 m_\tau^3 } {16\pi}\right)\tau_\tau f_{{h}}^2 |V_{\alpha\beta}|^2  
                       \left(1 - x\right)^2                                                                \nonumber \\         
             &    &    \times \left[ 
                                     1 + \frac{2\alpha}{\pi} {\mathrm{ln}} 
                                     \left( \frac {\mZ} {m_\tau} \right)+\cdots 
                       \right] \left[ 1 - \sin^2\theta  \right]  \nonumber  \\                                                                                                                                             
             &    &    \times  
                       \left[ 1 - y \left( \frac{2+x-y} {1-x} \right)                          
                            \sqrt{ 1 - y \frac{(2+2x-y)}{(1-x)^2} } \right]
         \label{equ:bhad}
\end{eqnarray}
where 
$x=m_{{h}}^2 / m_\tau^2$,
$m_{{h}}$ is the hadron mass, 
$f_{{h}}$ are the hadronic form factors, 
and $V_{\alpha\beta}$ are the CKM matrix elements, 
$V_{{ud}}$ and $V_{{us}}$,
for $\pi^-$ and $\kn$ respectively.
From an analysis of
$\pi^-\rightarrow\mu^-\bar{\nu}_\mu$ and
$\kn\rightarrow\mu^-\bar{\nu}_\mu$ decays, 
one obtains
$f_\pi |V_{{ud}}| = \FPIVUDval$ and
$f_{{K}} |V_{{us}}| = \FKVUSval$\cite[and references therein]{MARCIANO92A}. 
The ellipsis represents terms, estimated to be ${\cal{O}}(\pm 0.01)$\cite{MARCIANO92A},        
which are neither explicitly treated nor implicitly absorbed into $\GF$,
$f_\pi |V_{{ud}}|$, or $f_{{K}} |V_{{us}}|$.
The second term in brackets describes mixing with a fourth generation neutrino
while the third parametrises the effects of 
a non-zero $m_{\nu_3}$~\cite{MNUTAU}.

The fourth generation neutrino mixing affects all the tau branching
fractions with a common factor whereas a non-zero tau neutrino 
mass affects all channels with different kinematic factors.
Therefore, given sufficient experimental precision, these two effects 
could in principle be separated.

\section{Results}

We use the recently updated world average values for 
the measured tau branching fractions~\cite{LI97A}:
\begin{eqnarray}
  \Be  &  =  & \BRTEval; \\    
  \Bm  &  =  & \BRTMval; \\                      
  \Bp  &  =  & \BRTPval; \\    
  \Bk  &  =  & \BRTKval. 
\end{eqnarray}
Substituting in equations \ref{equ:blept} and \ref{equ:bhad} for 
the measured quantities we find that both 
$m_{\nu_\tau}$ and $\sin^2\theta$ are consistent with zero.

We therefore derive constraints on $m_{\nu_\tau}$ and $\sin^2\theta$
from a combined likelihood fit to the four tau decay channels. 
The likelihood is constructed numerically following the 
procedure of Ref.~\cite{NIM_LIKELIHOOD_PAPER} by randomly 
sampling all the quantities used according to their errors.
The CLEO measurement of the $\tau$ mass was used to further constrain $m_{\nu_3}$.
From an analysis of 
$\tau^+\tau^-$ 
$\rightarrow$ 
$(\pi^+n\pi^0\bar{\nu}_\tau)$
$(\pi^-m\pi^0\nu_\tau)$ 
events (with $n\leq2, m\leq2, 1\leq n+m\leq3$), CLEO determined the $\tau$ mass to be 
$m_\tau = (1777.8 \pm 0.7 \pm 1.7) + [m_{\nu_3}({\mathrm{MeV}})]^2/1400$ MeV\cite{CLEOWEINSTEIN}.
The likelihood for the CLEO and BES measurements to agree, as a function of 
$m_{\nu_3}$ is included in the global likelihood. 

The fit yields upper limits of 
\begin{eqnarray}
 m_{\nu_3}     & < & \MNUTvalB\,{\mathrm{MeV}}  \\  
 \sin^2\theta  & < & \FMIXvalB                   
\end{eqnarray}
at the 95\% C.L. or 
\begin{eqnarray}
 m_{\nu_3}     & < & \MNUTvalA\,{\mathrm{MeV}}  \\ 
 \sin^2\theta  & < & \FMIXvalA                   
\end{eqnarray}
at the 90\% C.L. 
These results improve on our previous determinations 
of $m_{\nu_3} < 42$\,MeV and $\sin^2\theta < 0.014$
at 90\% C.L.~\cite{MNUTAU}

\section{Discussion}

The limit on $m_{\nu_3}$ can be reasonably interpreted as a 
limit on $m_{\nu_\tau}$, since the mixing of 
$m_{\nu_3}$ with lighter neutrinos is also small~\cite{PDG96SHORT}.
The best direct experimental constraint on the tau neutrino mass 
is $m_{\nu_\tau} < 18.2$\,MeV at the 95\% confidence level\cite{TAU96_ALEPHNUTAU} 
which was obtained using many-body hadronic decays of the $\tau$.
While our constraint is less stringent, it is statistically independent.
Moreover, it is insensitive to fortuitous or pathological events 
close to the kinematic limits, details of the resonant structure 
of multi-hadron $\tau$ decays, and
the absolute energy scale of the detectors.
Since LEP has completed running on the Z it is unlikely that 
significantly improved constraints on $m_{\nu_\tau}$, using
multi-hadron final states, will be forthcoming
in the foreseeable future.

Future improved measurements of the tau branching fractions,
lifetime, and the tau mass from direct reconstruction 
would enable significant improvements to be made in the 
determinations of both $m_{\nu_\tau}$ and $\sin^2\theta$.
If CLEO and the b-factory experiments were to reduce the uncertainties 
on the experimental quantities by a factor of approximately 2, 
then the constraints on $m_{\nu_\tau}$ from the technique 
we have described would become the most competitive.
Were a tau-charm factory to be built, then the determination
of $m_{\nu_\tau}$ by direct reconstruction would again become
the most sensitive technique.  

Our upper limit on $\sin^2\theta$ is already the most stringent 
experimental constraint on mixing of the third and fourth 
neutrino generations.
  
\section*{Acknowledgements}
We would like to thank the Department of Physics,
Universidad Nacional de La Plata for their generous hospitality 
and the National Science Foundation for financial support.
J.S. gratefully acknowledges the support of the International
Centre for Theoretical Physics, Trieste.




\end{document}